\documentclass[aps,twocolumn,groupedaddress,floatfix,showpacs]{revtex4}

\usepackage{epsfig}
\usepackage{amssymb}

\begin{document}

\title{Nonadiabatic transitions between adiabatic surfaces: phase diffusion in superconducting atomic point contacts}

\author{H. Fritz$^1$ and J. Ankerhold$^2$}
\affiliation{$^1$ Physikalisches Institut, Albert-Ludwigs- Universit\"at Freiburg, 79104 Freiburg, Germany\\
$^2$Institut f\"ur Theoretische Physik, Universit\"at Ulm, Albert-Einstein-Allee 11, 89069 Ulm, Germany}

\date{\today}

\begin{abstract}
Motivated by experiments with current biased superconducting atomic point contacts the general problem of nonadiabatic transitions between adiabatic surfaces in presence of strong dissipation is studied. For a single channel device the supercurrent is determined by the diffusive motion of the superconducting phase difference on two Andreev levels. These surfaces are uncoupled only in the adiabatic limit of low to moderate transmissions, while for high transmissions curve crossings are important. Starting from a general master equation of the full density matrix an approximate time evolution equation for the populations on the adiabatic surfaces in the overdamped limit is derived from which the relevant observables can be obtained. Specific results for the case of atomic point contacts are in agreement with experimental observations that cannot be explained by conventional theory.
\end{abstract}


\maketitle

\section{Introduction}
In recent years curve crossing problems regained considerable attention in contexts such as quantum information processing, superconducting mesoscopic circuits, or wave packet dynamics in molecular structures.
The paradigmatic situation has been formulated by Landau, Zener, and St\"uckelberg \cite{landau,zener,stuckel} already during the heyday of quantum mechanics: the energy difference between the diabatic energy levels of two states, say $|1\rangle$ and $|2\rangle$, changes linearly in time as $v t$ with constant velocity $v$ through a crossing region while transitions between these states with frequency $\Delta_0/\hbar$ occur. The probability for staying in state $|1\rangle$ in the infinite future when starting  in $|1\rangle$ in the infinite past is given by the famous Landau-Zener-St\"uckelberg formula $P=\exp(- g)$ where the parameter $g=\pi \Delta_0^2/(\hbar v)$ controls whether the dynamics happens to be adiabatically ($g\gg 1$) or diabatically ($g\ll 1$).
Various modifications of this original setting have been studied, in particular, situations where the diabatic energy levels vary due to the dynamics of an intrinsic degree of freedom. For condensed phase systems this degree of freedom typically interacts with a dissipative environment so that due to its diffusive dynamics the crossing region (the so-called Landau-Zener range) is not traversed ballistically but rather stochastically. One prominent example is the charge transfer between donor and acceptor states in molecular structures, a process which is of relevance for molecular electronics \cite{nitzan}. Here, in most cases environmental degrees of freedom are fast compared to electronic transitions meaning that one is close to the diabatic limit. Important observables are then transfer rates between diabatic surfaces and time dependent populations on these surfaces. Since in general exact analytical solutions are impossible, approximate treatments have been developed e.g.\ based on master equations \cite{garg,hanggi,lehle,pollak}.

Much less attention has been paid to the diffusive dynamics on adiabatic surfaces and the impact of non-adiabatic transitions. One realization of this situation can be found in mesoscopic physics, namely, in the charge transfer through atomic size contacts between superconducting leads \cite{scheer,scheer2,yeyati}.
In fact, this device extends Josephson weak links \cite{barone} to junctions with variable transmissions even close to 1 with the relevant degree of freedom being again the phase difference between the superconducting reservoirs.
Theory shows \cite{beenakker} that for an isolated contact with one channel the Cooper pair current is carried by two Andreev bound states the energy levels of which depend on the phase difference and the transmission probability $\tau$ of the channel. For perfect transmission ($\tau=1$) the minimal gap between the levels closes, while it tends  towards the superconducting gaps in the reservoirs for $\tau\to 0$.

For contacts embedded in an electrical circuit the situation is more complex. First, the phase interacts with the electromagnetic modes of the surrounding and its dynamics becomes diffusive. Second, the contact can be externally driven by a current or a voltage bias. The charge transfer is then a combination of Cooper pair tunneling, responsible for a supercurrent peak, and multiple Andreev reflections (MARs) \cite{tinkham,averin,cuevas}, responsible for quasi-particle transfer leading to a rich subgap structure. The first process dominates for low voltages, the second one for higher voltages, while in a crossover region a subtle interplay of both processes appears \cite{chauvin}.

In this paper we concentrate on the current biased situation. Then, the phase diffuses on tilted Andreev levels which constitute the adiabatic levels of the system. Due to the small capacitance of the contact and the large admittance of the circuit the Brownian motion of the corresponding fictitious particle is overdamped so that this adiabatic approximation is justified. The theoretical framework to calculate averages of supercurrent and voltage, respectively, follows that of overdamped Josepshon junctions according to Ambegaokar, Halperin \cite{halperin} and Ivanchenko, Zil'bermann \cite{ivanchenko}. Ac-driven atomic point contacts have been treated in the same way in \cite{yeyati2}. Experimentally, this picture has been verified e.g.\ by measuring the supercurrent peak and its maximum (switching current) as functions of temperature \cite{goffman,cron}. Problems arise, however, close to the ballistic limit ($\tau$ close to 1), when the minimal gap between the adiabatic levels becomes comparable to typical diffusion times. Consequently, nonadiabatic transitions enter the game. Note that these transitions can also be interpreted as the precursors of MARs. In \cite{goffman} the standard Landau-Zener-St\"uckelberg formula has been applied to correct for those transitions, but failed to capture the experimental data. In this paper we provide a consistent theory which accounts for the overdamped Brownian motion on coupled adiabatic surfaces. Such a theory may in turn be of relevance in other contexts as well and particularly allows to calculate steady state currents and diffusive transition rates between {\em adiabatic} surfaces. For atomic point contacts its only limitation is that MARs are not included explicitly in the theory so that the full crossover between supercurrent peak and the subgap structure cannot be described. For this case
 (and a voltage biased contact) an effective approach has been outlined recently in \cite{chauvin}, which, however, is still somewhat {\em ad-hoc} as it combines a microscopic description of the MARs without environmental modes and the phase dynamics in presence of the circuit modes.

 The article is organized as follows: In Sec. \ref{adiabatic} we briefly describe the standard adiabatic model and introduce the relevant notation. A general approach to capture nonadiabatic transitions during diffusive motion on adiabatic surfaces is formulated in Secs.\ \ref{tls} and \ref{adiabatictls}, which is specialized to overdamped Brownian motion in \ref{overdamped}. There also an explicit expression for transition rates is derived. Results for the case of atomic point contacts are discussed in \ref{apc}. The role of quantum fluctuations is addressed in Sec.~\ref{quant}.

\section{Adiabatic phase dynamics}\label{adiabatic}

\begin{figure}
\begin{center}
\includegraphics[width=8.5cm]{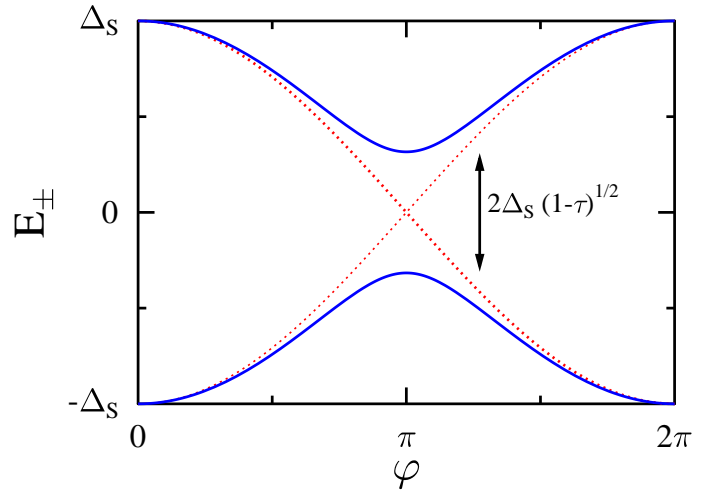}
\end{center}
\vspace*{-0.5cm}
\caption{\label{fig:andreevsurfaces}Energy surfaces $E_\pm$ of the bound Andreev states for a single channel with transmission $\tau$ (solid) together with the diabatic surfaces $V_{1,2}$ (dotted). The minimal energy gap between the Andreev levels at the Landau-Zener point $\varphi=\pi$ is indicated by the arrow. See text for details.}
\end{figure}

According to the conventional theory of a current biased point contact with superconducting leads \cite{goffman,yeyati}
transport is described by the phase dynamics on potential surfaces corresponding to energy surfaces of Andreev bound states.
For a single channel with two Andreev states the latter ones read
\begin{equation}
	E_{\pm}(\varphi)=\pm \Delta_S \sqrt{1 - \tau \sin^2(\varphi/2)}\,
	\label{eigenenergien}
\end{equation}
with $\Delta_S$ being the superconducting gap, $\tau$ the transmission probability through the contact, and $\varphi$ the phase difference between the superconducting reservoirs in the leads (see fig.~\ref{fig:andreevsurfaces}).
In the basis of right and left moving waves (Bogoliubov-de Gennes theory) the underlying Hamiltonian of this two level system is given by \cite{ivanov}
\begin{equation}
	H_A= \Delta_S \left(
		\begin{array}{cc}
			\cos(\varphi/2) & \sqrt{1-\tau} \sin(\varphi/2) \\
			\sqrt{1 - \tau} \sin(\varphi/2) & -\cos(\varphi/2)
		\end{array}
	\right)\, ,
	\label{andreev_hamiltonian}
\end{equation}
 so that transitions between ''diabatic'' surfaces $V_{1/2}=\pm \cos(\varphi/2)$ are mediated by off-diagonal couplings
 which are maximal at $\varphi=\pi$  corresponding  to a minimal energy gap $2\Delta_S \sqrt{1-\tau}$  between the adiabatic surfaces (\ref{eigenenergien}). This gap closes in the ballistic limit $\tau=1$.
The current carried by each of the surfaces (\ref{eigenenergien}) follows from
\begin{equation}
	I_\pm(\varphi) = \frac{1}{\varphi_0} \frac{\partial E_\pm(\varphi)}{\partial \varphi}
				= \mp \frac{e \tau \Delta_S}{2 \hbar} \frac{\sin \varphi}{\sqrt{1- \tau \sin^2(\varphi/2)}}
				\quad
	\label{andreev_strom}
\end{equation}
with the reduced flux quantum $\varphi_0=\hbar/2e$.

\begin{figure}
\begin{center}
\includegraphics[width=7.5cm]{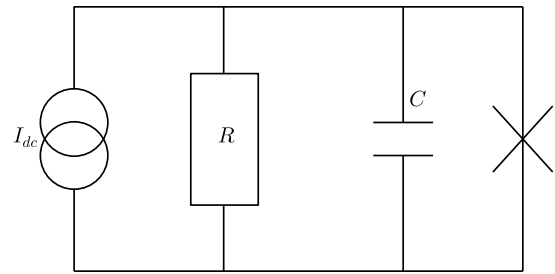}
\end{center}
\vspace*{-0.5cm}
\caption{\label{fig:circuit}Equivalent circuit of the resistively and capacitively shunted junction (RCSJ) model including a weak link.}
\end{figure}
A realistic description of transport across the contact has also to account for its electromagnetic environment. According to the RCSJ model one writes for the circuit in fig.~\ref{fig:circuit}
\begin{equation}
	I_{dc}+I_{n} = \varphi_0 C \frac{d^2 \varphi}{dt^2} + \varphi_0 \frac{1}{R}
			 \frac{d \varphi}{dt}	+ I(\varphi)\, ,
	\label{stromkreis_3}
\end{equation}
where an applied bias current consists of a dc-component $I_{dc}$ and a fluctuating component $I_{n}$ which
obeys Johnson-Nyquist characteristics, i.e.,
\begin{equation}
	\left\langle I_n(t) \right\rangle =0\ , \ \left\langle I_n(t) I_n(0) \right\rangle = \frac{2k_BT}{R}\delta(t)\, .
\end{equation}
Further, $C$ denotes the capacitance, $R$ the resistance, and $I(\varphi)$ the supercurrent through the contact. This way, the phase dynamics is equivalent to the Brownian motion of a fictitious particle with mass $m=\varphi_0^2 C$ and friction constant $\gamma=1/R C$. In the adiabatic approximation the motion of the phase is assumed to be much slower than any other relevant process meaning that the supercurrent $I(\varphi)$ is simply the Boltzmann weighted sum of the individual currents $I_\pm$, i.e.,
\begin{equation}
I(\varphi)=I_-(\varphi)\, {\rm tanh}\left[\beta E_+(\varphi)\right]
\end{equation}
with inverse temperature $\beta=1/k_{\rm B} T$. The effective potential felt by the fictitious particle is then
\begin{equation}
U(\varphi)=\varphi_0\left[-I_{dc}\, \varphi+\int_0^\varphi d\varphi' I(\varphi')\right]\, .
\label{meanpot}
\end{equation}
In the actual experimental set-up the capacitance is negligible leading to strongly overdamped phase dynamics. Hence, the adiabatic approximation is indeed justified as long as the adiabatic surfaces are sufficiently separated from each other (a more detailed condition will be given below).

To calculate current and voltage across the contact it is much more convenient to work with the probability distribution $W(\dot{\varphi},\varphi)$ in phase-space corresponding to the Langevin dynamics (\ref{stromkreis_3}). In the overdamped domain the only relevant information is carried by the marginal probability distribution $n(\varphi,t)=\int d\dot{\varphi}\, W(\dot{\varphi},\varphi)$ and the Fokker-Planck equation for the full distribution reduces to a Smoluchowski equation of the form
\begin{equation}
	\frac{\partial n(\varphi,t)}{\partial t} = \frac{R}{\varphi_0^2}\frac{\partial}{\partial \varphi}
		\left[U'(\varphi)+ k_{\rm B} T \frac{\partial}{\partial \varphi}  \right] n(\varphi,t)\, .
	\label{smoluch_strom}
\end{equation}
with $U'(\varphi)=dU(\varphi)/d\varphi$.
Now, for fixed applied bias current $I_{dc}$ the dynamics tends towards a steady state distribution $n_{st}(\varphi)$ for longer times from which the average supercurrent through the junction $I_J(I_{dc})=\langle I(\varphi)\rangle_{st}$ and the mean value of the voltage across the contact $U(I_{\rm dc})=\varphi_0 \langle \dot{\varphi}\rangle_{st}$ can be calculated. Upon varying $I_{dc}$ one gains the supercurrent voltage characteristics $I_J(U)$.

\section{Dissipative two level system}\label{tls}
The adiabatic treatment is based on a separation of time scales, which may become critical in certain ranges of parameter space. The goal of the next two sections is to derive  formally exact equations of motion for the density matrix in the adiabatic basis of the two level system thus including nonadiabatic transitions between them. This in turn allows to systematically go beyond the simple approach presented above. The corresponding analysis is completely general and applies to all situations, where diffusive dynamics on single adiabatic surfaces tends to break down. Accordingly, we slightly change notation for the relevant degree of freedom from $\varphi$ to $q$  to stress the close analogy to the dynamics of a fictitious particle.

The starting point is a standard system + reservoir Hamiltonian of the form $H = H_0 + H_W + H_B$, where
\begin{equation}
	H_0 = \frac{p^2}{2m}\openone+ \left(			
				\begin{array}{cc}
						V_1(q)		& \Delta_0(q)/2 \\
						\Delta_0(q)/2	& V_2(q)
				\end{array}
			\right)  ,
	\label{H_WFP}
\end{equation}
describes the dynamics on two coupled diabatic surfaces $V_{1,2}$ which cross at the Landau-Zener (LZ)-point $q^*$, i.e.\ $V_1(q^*)=V_2(q^*)$,  but are separated sufficiently away from $q^*$ by energies larger than the thermal energy scale $k_{\rm B} T$.
Further,
\begin{equation}
	H_W = \left(- q \sum_{n=1}^N {c_n x_n} + q^2 \sum_{n=1}^N \frac{c_n^2}{2m_n \omega_n^2}\right)\openone. 	
	\label{wechselwirkungsanteil}
\end{equation}
denotes the interaction of the relevant dynamical degree of freedom to a heat bath
\begin{equation}
	H_B = \sum_{n=1}^{N}{\frac{p_n^2}{2m_n}} + \frac{1}{2} \sum_{n=1}^{N}{m_n \omega_n^2 x_n^2}
	\label{bad_hamilton}
\end{equation}
consisting of a large number of harmonic degrees of freedom. In the continuum limit the spectral density of this oscillator bath is chosen to be ohmic below a cut-off frequency $\omega_c$, i.e.\ $J(\omega)=m \gamma\omega$ with friction strength $\gamma$, and zero otherwise.
The dynamics of the reduced density matrix $\rho(t)={\rm Tr}_B\{\rho_{\rm tot}(t)\}$ can be represented formally exactly in terms of path integrals \cite{weiss}. In the domain of higher temperatures though, an explicit time evolution equation can be derived \cite{garg,breuer,weiss}, namely,
\begin{equation}
	\frac{\partial \rho}{\partial t} = - \frac{i}{\hbar} \left[H_0, \rho \right]
							-\frac{m\gamma k_B T}{\hbar^2} \left[ q,\left[ q, \rho \right] \right]
							-\frac{i \gamma}{2\hbar}\left[ q,\left\{ p, \rho \right\} \right]\, ,
	\label{operatorgleichung}
\end{equation}
where  $\{\,,\,\}$ denotes the anti-commutator.
	The detailed conditions for the validity of this approximation are: $\hbar \beta \omega_c, \hbar \beta \gamma\ll 1$. Note that this regime particularly includes the domain of high friction $\gamma/\omega_0\gg 1$ at sufficiently elevated temperatures where $\omega_0$ is a typical system frequency scale.

For the two level problem the above master equation has been the starting point for various approximations \cite{weiss}. In particular, in the high temperature domain $\omega_0\hbar\beta\ll 1$ it is convenient to represent it in terms of the Wigner transform \cite{pollak}
\begin{equation}
	W(q,p,t) = \int \frac{dq'}{2\pi \hbar} \; {\rm e}^{-ipq'/\hbar}
					\left\langle \frac{q + q'}{2}\Big| \rho(t)  \Big|\frac{q - q'}{2} \right\rangle\, .
	\label{wigner_trafo}
\end{equation}	
This gives in leading order in $\hbar$ the Wigner-Fokker-Planck equation
\begin{eqnarray}
	\frac{\partial W}{\partial t}  &=& - \frac{p}{m} \frac{\partial W}{\partial q}
					+  V'_+
							 \frac{\partial W}{\partial p}
	 +\gamma\left(\frac{\partial}{\partial p} p W\right. \nonumber\\
&&\left. +m k_BT \frac{\partial^2 W}{\partial p^2}\right)
	 + \frac{V'_-}{4}
				 \frac{\partial}{\partial p}\left\{ \sigma_z, W \right\}\nonumber\\
				&&- \frac{i}{2\hbar} \left[ V_- \sigma_z
																	+  \Delta_0 \sigma_x, W \right]\, ,
	\label{WFP}
\end{eqnarray}
where we introduced the sum and the difference of the diabatic surfaces, respectively,
\begin{equation}
V_+=(V_1 + V_2)/2\ \ , \ \  V_-=V_1 - V_2\,.
\end{equation}
For practical applications, however, this set of time evolution equations for the matrix elements $W_{ij}(q,p,t), i,j=1,2$ is of limited use only. Namely, it turns out that singularities appear for the off-diagonal Fokker-Planck operators due to highly oscillatory terms $V_-/\hbar$, particularly in cases where the so-called reorganisation energy $E_r$ is large compared to $k_{\rm B} T$, $\hbar\omega_0$ \cite{pollak}. Here, $E_r=|V_-(q_{\rm min})|$ is the energy needed to switch from the minimum  $q_{\rm min}$ of the lower diabatic surface to the higher lying one. In case of the two Andreev levels introduced in the previous section, we have $E_r\approx\Delta_S$. The strategy to proceed is then to use approximate solutions for the off-diagonal elements in the limit $k_{\rm B} T, \hbar\omega_0\ll E_r$ to derive a set of two coupled effective equations for the diagonal densities \cite{zusman,pollak,lehle}. Coupling terms describe the impact of adiabatic transitions between the diabatic surfaces the strength of which is measured by a LZ type of factor $\Delta_0^2/|V_-'|$.
This way, populations and rate constants for the electron transfer between donor and acceptor states in  molecular structures have been calculated numerically and compared to analytical findings, see e.g.\ \cite{pollak,lehle}. The diabatic surfaces in this situation are assumed to be harmonic so that the system, starting initially from the donor state, relaxes towards thermal equilibrium in the long time limit.

In the present problem the situation is different though: here, the relevant physical observables are determined as expectation values on {\em adiabatic} surfaces. To calculate them from the diabatic representation is possible in principle, but fails in practice since it requires the accurate knowledge of the full diabatic density matrix, which is not available as discussed above. In particular, in a steady state off-diagonal elements typically do not vanish. Hence, we follow another route and first transform  the full master equation (\ref{WFP}) to the adiabatic basis before further approximations are applied.

\section{Dynamics of the density matrix in the adiabatic basis}\label{adiabatictls}
The master equation (\ref{WFP}) describes the dynamics of the Wigner distribution in the so-called diabatic basis, in which the Hamiltonian becomes diagonal for the artificial spin held fixed, that is for vanishing coupling $\Delta_0$. In this section the corresponding master equation in the adiabatic basis is derived, the latter one obtained by diagonalizing   $H_0'=H_0-(p^2/2m)\, \openone$, i.e.,
\begin{equation}
	H'_0 =
		\left(
			\begin{array}{cc}
				V_1	 & \Delta_0/2 \\
				\Delta_0/2 & V_2
			\end{array}
		\right)\, .
		\label{H_0_strich}
\end{equation}
The corresponding unitary transformation is given by
\begin{equation}
	U (\phi)=
		\left(			
			\begin{array}{cc}
				\cos \phi & \sin \phi \\
				-\sin \phi & \cos \phi
			\end{array}
		\right)\, ,
	\label{drehmatrix}
\end{equation}
where $\phi(q) = (1/2) \arctan\left[\Delta_0/V_-(q)\right]$.  One finds for $ H'_{ad}=	UH'_0U^\dagger$ the expected result
\begin{equation}
	H'_{ad}=
		\left(
			\begin{array}{cc}
				V_+ + V_{ad}	 & 0 \\
				0 & V_+ - V_{ad}
			\end{array}
		\right)
	\label{diagonal_H_0}
\end{equation}
with the adiabatic surfaces $V_+\pm V_{ad}$ where
\begin{equation}
V_{ad}=\frac{1}{2}\sqrt{V_-^2 + \Delta_0^2}\,
\end{equation}
describes the gap between them.

To transform the full Hamiltonian $H_0$, the unitary operator must also be applied to the kinetic part. From
$U\, p\, U^\dagger = p - \hbar k(q) \sigma_y$ with
\begin{equation}
	k(q) = \frac{\partial \phi(q) }{\partial q}
				= \frac{\Delta'_0 V_-(q) - \Delta_0 V_-'(q)}{8\, V_{ad}(q)^2}
	\label{k_faktor}
\end{equation}
 we obtain in the adiabatic representation for $H_{ad}= U H_0 U^\dagger$
\begin{eqnarray}
		\lefteqn{H_{ad}= H'_{ad}}\nonumber\\
&&  +\openone\left[ \frac{p^2}{2m}
							+ \frac{\hbar^2 k(q)^2 }{2m}- \frac{\hbar}{2m} \sigma_y \{p, k(q)\}\right]\, .
	\label{H_ad}
\end{eqnarray}
In the adiabatic approximation discussed in the first section, the last two terms are neglected. Here, we retain them  to find together with $U(H_W + H_B)U^\dagger = H_W + H_B$ the exact form of the master equation (\ref{operatorgleichung}) for the density matrix in the adiabatic basis
$\rho_{ad}= U \rho U^\dagger$ as
\begin{eqnarray}
	\frac{\partial \rho_{ad}}{\partial t} &=& - \frac{i}{\hbar}[H_{ad}, \rho_{ad}]
												- \frac{m\gamma k_B T}{\hbar^2} [q,[q,\rho_{ad}]]
												\nonumber\\
												&&- \frac{i \gamma}{2  \hbar} [q,\{p,\rho_{ad}\}]+ \frac{i \gamma}{2 } [q,\{k(q)\sigma_y,\rho_{ad}\}]\, .
	\label{adiab_op_gleichung}
\end{eqnarray}
Obviously, the last term accounts for nonadiabatic transitions in the adiabatic dynamics. Its role becomes much more transparent in the Wigner transformed master equation, where we have
\begin{eqnarray}
	\frac{\partial W_{++}}{\partial t}  &=& {\cal L}_+ W_{++}+ k(q) \frac{p}{m} \left( W_{+-} + W_{-+}\right)
						\nonumber \\
	\frac{\partial W_{--}}{\partial t} &=& {\cal L}_- W_{--} - k(q) \frac{p}{m} \left( W_{+-} + W_{-+} \right)
						\nonumber \\
	\frac{\partial W_{+-}}{\partial t}  &=& {\cal L}_0 W_{+-} - k(q) \frac{p}{m} \left(W_{++} - W_{--}\right) \nonumber\\	 &&-\frac{2i}{\hbar} V_{ad} W_{+-} \label{adia_WFP}
\end{eqnarray}
and likewise for the complex conjugate $W_{-+}=W_{+-}^*$ with the operators
\begin{equation}
\label{lop}
{\cal L}_\eta=-\frac{p}{m} \frac{\partial}{\partial q}
						+\left( V_+' +\eta\,  V'_{ad} \right) \frac{\partial}{\partial p}
						+ \gamma  \left( \frac{\partial}{\partial p} p
						 + m k_B T \frac{\partial^2}{\partial p^2}\right)\,
\end{equation}
where $\eta=0,+,-$. Transitions between the adiabatic surfaces $V_++V_{ad}$ and $V_+-V_{ad}$ occur via transitions to off-diagonal elements. The corresponding coupling strength is given by  $k(q) p/m$ which contains both the {\em inverse} of a LZ type of factor $k(q)$ measuring the differences between the diabatic forces compared to the adiabatic energy gap and a dynamical factor $p/m$. Away from the LZ point $q^*$ one has $|V_-|\gg \Delta_0$ so that the first factor is small and together with a slow motion in $p$ nonadiabatic transitions are negligible. In the LZ range around $V_-=0$, however, the crucial quantity is $|V_-'/\Delta_0|$, which may become large. For instance, for an atomic point contact the LZ range is located around $\varphi=\pi$ where $|V_-'/\Delta_0|\propto 1/\sqrt{1-\tau}$. As a consequence, the adiabatic approximation breaks down for $\tau \to 1$ and the dynamics of the density matrix follows from the full set of equations only. As in case of the diabatic equations (\ref{WFP}),  a direct numerical evaluation of (\ref{adia_WFP}) is in most cases prohibitive though. Namely, the off-diagonal elements $W_\pm$  tend to oscillate strongly for increasing $V_{ad}$, i.e. away from the LZ-domain. Physically, this reflects the fact that these off-diagonal elements are relevant only in a domain around the LZ point $V_-=0$, while outside they are effectively washed out. The idea to proceed in the overdamped limit is thus similar as in the diabatic representation: one formally solves for the localized dynamics of $W_{+-}$ and $W_{-+}$ and inserts these results into the equations for the diagonal elements. Eventually one arrives at two coupled effective equations of motion for the latter ones which are amenable to numerical approaches. Corresponding approximations are then adapted to the diffusive dynamics on the {\em adiabatic} surfaces.

\section{Overdamped dynamics}\label{overdamped}
In the regime of strong friction the dynamics in position is slow, while equilibration in momentum occurs on the fast time scale $1/\gamma$. This separation of time scales allows for an explicit elimination of the off-diagonal elements in (\ref{adia_WFP}) to obtain a set of equations of motion for the populations alone coupled by an effective position dependent transition factor.

\subsection{Population dynamics}
We start by writing $\bar{W}_{+-}=\exp(2i V_{ad} t/\hbar) W_{+-}$ so that  for the off-diagonal elements in (\ref{adia_WFP}) one has
\begin{eqnarray}
\frac{\partial \bar{W}_{+-}}{\partial t}&=&{\cal L}_0\bar{W}_{+-}
						-  \frac{p}{m} k(q) \exp\left(\frac{2i}{\hbar} V_{ad} t \right) (W_{++} - W_{--})\nonumber\\
						&&+\frac{p}{m} \frac{2it}{\hbar} V'_{ad}  \bar{W}_{+-}.
	\label{korr_WFP}
\end{eqnarray}
Now, the propagator of the bare system obeying $d{\cal G}/dt={\cal L}_0 {\cal G}$ is to be calculated in the overdamped limit $\gamma/\omega_0\gg 1$ with $\omega_0$ being a typical system frequency. Accordingly, we look for times within the window $1/\gamma<t<\gamma/\omega_0^2$, where the upper bound follows from the fact that off-diagonal elements are determined by the dynamics on a much shorter time scale $\hbar/\Delta_0\ll \gamma/\omega_0^2$ (see below). This way, one finds (see appendix)
\begin{equation}
	{\cal G}(q,p;\bar{q},\bar{p};t) = \delta(q-\bar{q}) \frac{1}{\sqrt{2\pi m k_B T}}\,
			{\rm e}^{-[ p - p_Q(t)]^2/2m k_B T}\, .
	\label{greensfkt_approx}
\end{equation}
Further, the order of magnitude of the last term in (\ref{korr_WFP}) can be estimated in the LZ range as $p V'_{ad} t/m\sim \langle (p_Q/m) V'_{ad}\rangle_\beta (\hbar/\Delta_0)$, which can thus be neglected if
\begin{equation}
	\left| \frac{k_B T}{\Delta_0} \frac{ \omega_0}{\gamma}\right|
	\ll 1\, .
	\label{bed_korr}
\end{equation}
Note that for $\hbar\omega_0\beta\ll 1$ the above condition is stronger than $\hbar/\Delta_0\ll \gamma/\omega_0^2$ and thus the relevant one.
Now, solving (\ref{korr_WFP}) formally and plugging the result into the equations of motion (\ref{adia_WFP}) for the diagonal elements one arrives at
\begin{eqnarray}
	\frac{\partial W_{++}}{\partial t}
				&=& {\cal L}_{+} W_{++}
						 + 2\, \mbox{Re} \int_0^t\!\!\! d\bar{t}\!\!\int d\bar{q}d\bar{p}\,{\cal G}(q,p;\bar{q},\bar{p};t-\bar{t}\,)
						\nonumber \\	
				&&		\times 	\frac{p\bar{p}}{m^2} k(q)k(\bar{q})
						\, 	{\rm e}^{-{2i}[V_{ad}(q) t -V_{ad}(\bar{q})\bar{t}\,]/\hbar}\nonumber\\
				&&\times			[W_{--}-W_{++}](\bar{q},\bar{p},\bar{t}\,)\, .
			\label{diagonaleff}
\end{eqnarray}
and likewise for $W_{--}$. In the overdamped limit the Wigner distributions factorize according to
\begin{equation}
	W_{\pm}(q,p,t)=\frac{1}{\sqrt{2\pi m k_B T}} \exp \left( - \frac{p^2}{2m k_B T} \right)
								n_{\pm}(q,t)
	\label{annahme_WSG}
\end{equation}
so that the marginal distributions in position
\begin{equation}
n_{\pm}(q,t)=\int dp W_{\pm \pm}(p,q,t)
\end{equation}
 readily follow together with (\ref{greensfkt_approx}) from (\ref{diagonaleff}). We thus gain the central result of this work, namely, a time evolution equation for the adiabatic populations including non-adiabatic transitions, i.e.,
\begin{eqnarray}
	\frac{\partial  n_+}{\partial t} &=& \tilde{{\cal L}}_+ n_+ +
			\frac{k(q)^2}{m^2} \widetilde{C}_{pp}^>\left(\frac{2 V_{ad}}{\hbar}\right)(n_{-} - n_{+})
	\nonumber \\	
	\frac{\partial n_-}{\partial t}  &=& \tilde{{\cal L}}_- n_{-} -
			\frac{k(q)^2}{m^2} \widetilde{C}_{pp}^> \left(\frac{2 V_{ad}}{\hbar}\right)(n_{-} - n_{+})\, .
	\label{adiab_WSG}
\end{eqnarray}
with the Smoluchowski operators
\begin{equation}
\tilde{{\cal L}}_\eta=\frac{1}{m\gamma}\frac{\partial}{\partial q}\left(V_+'+\eta V'_{ad}+\frac{1}{\beta}\frac{\partial}{\partial q}\right)
\end{equation}
and $\eta=+,-$.
Here, $\tilde{C}_{pp}^>(\omega)$ denotes the Fourier transform  of the momentum-momentum correlation function $C_{pp}(t)=\langle p(t) p(0)\rangle$ of a local harmonic oscillator given by \cite{weiss}
\begin{eqnarray}
C_{pp}(t)&=&\frac{m^2\hbar}{2\pi}\int_{-\infty}^\infty d\omega \chi''(\omega)\omega^2\Big[{\rm coth}\left(\frac{\omega\hbar\beta}{2}\right)\cos(\omega t)\nonumber\\
&&-i\sin(\omega t)\Big]
\label{cpp}
\end{eqnarray}
in the regime of high temperatures where the real part dominates \cite{remark}. System information is carried by the dynamical susceptibility $\chi=\chi'+i \chi''$, the imaginary part of which  becomes in the strong friction limit independent of the local oscillator frequency
\begin{equation}
\label{chiom}
\chi''(\omega)=\frac{1}{m}\frac{\gamma\omega}{\omega^4+\omega^2\gamma^2}\,
\end{equation}
and
\begin{equation}
\tilde{C}_{pp}^>(\omega)=\frac{2 m^2}{\beta} \chi''(\omega)\omega\, .
\end{equation}
Thus, we see from (\ref{adiab_WSG}) that the nonadiabatic coupling between the adiabatic surfaces contains in addition to the Landau-Zener factor $k(q)$ dynamical information in terms of the power spectrum of the momentum correlations at the local transition frequency $\Omega=2V_{ad}(q)/\hbar$ between the adiabatic surfaces. Since the relevant domain for transitions is the LZ range where the gap between the adiabatic surfaces becomes small, one has $\Omega_0\equiv 2 V_{ad}(q^*)/\hbar\ll \gamma$ and in most cases also $\Omega_0\hbar\beta< 1$ which justifies the high temperature analysis given above. A simple extension of the result (\ref{adiab_WSG}) to lower temperatures ($\Omega\hbar\beta>1$) uses the full expression (\ref{cpp}) (see also next section).

\subsection{Transition rates}

In this section we derive a transparent expression for the transition rate between the adiabatic surfaces as defined by the loss of population in the steady state to the other surface  if initially only one, say the lower, surface is populated.
This analysis is independent of the previous one and exploits time-dependent perturbation theory. It thus applies only if the nonadiabatic coupling between the adiabatic surfaces remains sufficiently small. It turns out though that this approach provides direct insight into the non-adiabatic processes captured in (\ref{adiab_WSG}).

One writes
$H_{ad} = H_D + W_0$
with
\begin{equation}
	H_D=
	\left(
			\begin{array}{cc}
				H_+ & 0 \\
				0 & H_-
			\end{array}
	\right)
	=  V_{ad} \sigma_z + \openone\left(V_+ + \frac{p^2}{2m}
							+ \frac{\hbar^2 k(q)^2 }{2m} \right)
\end{equation}
and the ''perturbation''
\begin{equation}
	W_0=
	\left(
		\begin{array}{cc}
				0 & iW' \\
				-iW' & 0
		\end{array}
	\right)
	= - \frac{\hbar}{2m} \sigma_y \{p, k(q)\}.
\end{equation}
The propagator in the interaction picture is up to second order
\begin{eqnarray}
	G_I(t) &\approx& 1 -\frac{i}{\hbar} \int_{-\infty}^{t}{dt' \; W_I(t')}\nonumber\\
			&&-\frac{1}{\hbar^2}\int_{-\infty}^{t}{dt' \; \int_{-\infty}^{t'}{dt'' \; W_I(t')W_I(t'')}}\, .
	\label{neumann_reihe}
\end{eqnarray}
The initial density matrix is chosen as
\begin{equation}
	\rho_I(0) = \left| - \right\rangle \left\langle - \right|
				\otimes \bar{\rho}_-
				=\left(
						\begin{array}{cc}
							0 & 0 \\
							0 & \bar{\rho}_-
						\end{array}
				\right)
\end{equation}
with $|-\rangle$ being the eigenstate of the lower surface and $\rho_-(q)$ the corresponding initial distribution in $q$ so that the population  $P_-(t)={\rm Tr}\{|-\rangle\langle -|\, \rho_I(t)\}$ is given by
\begin{eqnarray}
	P_-(t) &\approx& 1 - \frac{2 }{\hbar^2}\mbox{Re} \int_{-\infty}^{t}\!\! dt'\int_{-\infty}^{t'} \!\!dt''
							\mbox{Tr}\left\{(G_-^\dagger W' G_+)(t')\right.\nonumber\\
&& \left.\times(G_+^\dagger W' G_-)(t'')\bar{\rho}_-  \right\}\, .
	\label{P_t_1}
\end{eqnarray}
Accordingly, the change in time is obtained to read
\begin{equation}
 \dot{P}_-(t) \approx-\frac{1}{\hbar^2} \int_{-\infty}^{+\infty} d\tau  \mbox{Tr}\left\{
							{\rm e}^{-iH_-\tau/\hbar} W' {\rm e}^{iH_+\tau/\hbar} W'
							\bar{\rho}_-\right\}\, ,
				\label{P_t_2}
\end{equation}
where one approximates
\begin{equation}
	\exp(iH_+\tau/\hbar) \approx \exp(iH_-\tau/\hbar)\exp(i 2V_{ad}(q)\tau/\hbar) \, .
\end{equation}
For sufficiently long times, but short compared to the relaxation time, we have $\dot{P}_-(t)\approx -\Gamma$ with the formal expression for the transition rate
\begin{equation}
	\Gamma= \frac{1}{\hbar^2} \int_{-\infty}^{+\infty} d\tau  \mbox{Tr}\left\{
							{\rm e}^{-iH_-\tau/\hbar} W' {\rm e}^{iH_-\tau/\hbar}
							{\rm e}^{i 2V_{ad}\tau/\hbar} W' \bar{\rho}_- \right\}\, .
\end{equation}
Now, using
\begin{equation}
	\exp(-iH_-\tau/\hbar)W'\exp(iH_-\tau/\hbar)
		\approx \frac{\hbar}{m} p(-\tau) k(q)
\end{equation}
one finds
\begin{eqnarray}
	\Gamma &\approx &  \int_{-\infty}^{+\infty}{d\tau  \mbox{Tr}\left\{
									\frac{k(q)}{m\hbar} p(-\tau) {\rm e}^{i 2V_{ad}\tau/\hbar} W' \bar{\rho}_- \right\}}
						\nonumber \\
					&\approx & \int_{-\infty}^{\infty} d\tau \; \mbox{Tr}\left\{
									 p(\tau)p(0)  {\rm e}^{- i2 V_{ad} \tau/\hbar} \frac{k(q)^2}{m^2}\bar{\rho}_- \right\}.
\end{eqnarray}
In the overdamped limit the total density matrix $\bar{\rho}$ factorizes in an equilibrium distribution for the momentum and a time dependent distribution in position. Hence, one arrives with (\ref{cpp}) at
\begin{eqnarray}
\Gamma &=&	 \int_{-\infty}^{\infty}d\tau \int dq'\;
	 C_{pp}(\tau)\,  {\rm e}^{-i2V_{ad}\tau/\hbar} \frac{k(q')^2}{m^2} n_{-}(q')
				\nonumber \\
	&=& \int dq\,  k(q)^2\,  \frac{4 V_{ad}^2}{\hbar }
						\left[\coth\left( \frac{V_{ad}}{k_B T} \right) - 1  \right]\nonumber\\
					&&\hspace*{0.75cm}\times\, {\chi}''\left( \frac{2 V_{ad}}{\hbar} \right)\  n_{-}(q)	\, .
			\label{st_rate}
\end{eqnarray}
As expected, the transition rate contains a transition factor appearing also in (\ref{adiab_WSG}), particularly, the Fourier transform of the momentum-momentum correlation for strong friction. Note however that the full set of equations remains valid also for larger couplings where the perturbative approach fails provided (\ref{bed_korr}) applies.

\section{Results for an atomic point contact}\label{apc}

With the general result (\ref{adiab_WSG}) at hand we now return to the system already addressed in the first section, namely, adiabatic dynamics of the phase in  single channel superconducting atomic point contacts. We then have for the populations on the adiabatic surfaces
\begin{eqnarray}
	\frac{\partial n_{+}(\varphi, t)}{\partial t} &=& {\cal L}_+ n_{+}(\varphi,t)
+{\cal K}(\varphi)( n_{-} - n_{+})(\varphi,t)
	 	\nonumber \\
	\frac{\partial n_{-}(\varphi, t)}{\partial t} &=& {\cal L}_- n_{-}(\varphi,t) - {\cal K}(\varphi) ( n_{-} - n_{+})(\varphi,t)
	 	\label{andreev_WSG}
\end{eqnarray}
where the individual diffusion operators read ${\cal L}_\pm=(R/\varphi_0^2 \beta)\partial_\varphi {\cal J}_{\pm}$ with the flux operators
\begin{equation}
{\cal J}_\pm= \frac{\partial}{\partial \varphi}
	 		+ \beta \left( \varphi_0 I_{dc}
	 			\mp \frac{\tau \Delta_S}{4}\frac{\sin \varphi}{\sqrt{1- \tau \sin^2(\varphi/2)}} \right)\, .
\end{equation}
The phase dependent coupling is
\begin{equation}
{\cal K}(\varphi)=\frac{R}{8 \varphi_0^2 \beta}
	 		\frac{1 -\tau}{\big[ 1 - \tau \sin^2(\varphi/2) \big]^2[1+(2RC V_{ad}/\hbar)^2]}\,
\end{equation}
and contains $V_{ad}=\Delta_S \sqrt{1-\tau \sin(\varphi/2)^2}$.

Before we proceed let us first estimate the range of  validity of the above equations of motion according to the condition (\ref{bed_korr}) with the minimal $\Delta_0(\varphi=\pi)=2 \Delta_S \sqrt{1-\tau}$. For typical experimental parameters  \cite{goffman,cron}  $\omega_0/\gamma\sim 0.04$, and $k_{\rm B}T/\Delta_S\sim 0.5$ one must obey $1-\tau>10^{-4}$ meaning that the approach only fails for transmissions extremely close to the full ballistic regime $\tau=1$. Note that in this latter limit ${\cal K}(\varphi=\pi)\propto 1/(1-\tau)$ diverges. The above restriction can also be understood from a different point of view. Namely, the approach discussed here has to assume that the instantaneous voltage $\varphi_0 \dot{\varphi}$ produced when the phase diffuses with finite velocity through the LZ range is sufficiently smaller than the superconducting gap $\Delta_S$ so that the massive production of quasi-particles is suppressed. The force experienced by the phase in the LZ domain is $V_{ad}'$ which for $\tau\to 1$ becomes large in the vicinity of $\varphi=\pi$ according to $(\pi-\varphi)/\sqrt{1-\tau}$.
Consequently, the width of the LZ range tends to vanish for $\tau\to 1$.
Thus, estimating the typical variance in phase to be on the order of the thermal length scale, one derives precisely the above condition. Hence, the present formalism is not able to capture the full crossover to larger voltages where multiple Andreev reflections (MARs) play the dominant role. However, it does reveal the impact of nonadiabatic transitions in the $I-U$ range where the maximum of the supercurrent peak is observed. This is discussed in detail below.

For this purpose, one calculates the mean supercurrent through the contact from the  steady-state populations
$n_\pm^{st}(\phi)$ obeying $n_\pm^{st}(0)=n_\pm^{st}(2\pi)$ according to
\begin{equation}
	\left\langle I_J \right\rangle (I_{dc}) = \frac{1}{\varphi_0} \int_0^{2\pi}{d\varphi \;
			V'_{ad}(\varphi)
			\big[ n_{+}^{st}(\varphi) - n_{-}^{st}(\varphi) \big]}\, .
			\label{numerik_strom}
\end{equation}
The corresponding voltage across the contact is proportional to the mean phase velocity $\left\langle U \right\rangle = \varphi_0 \langle \dot{\varphi} \rangle$ and one has
\begin{equation}
	\left\langle U \right\rangle (I_{dc})= - \frac{R}{\varphi_0 \beta}
					\int_0^{2\pi}{d\varphi \; \big[ {\cal J}_{+} n_{+}^{st}(\varphi)}
					+ {\cal J}_{-} n_{-}^{st}(\varphi) \big]\, .
				\label{numerik_spannung}
\end{equation}
The equations (\ref{andreev_WSG}) are now solved numerically with the thermal initial conditions $n_\pm(t=0)=\exp[-\beta (V_+\pm V_{ad})]/Z$ and for fixed values of the external bias current. Throughout the rest of the paper we use typical experimental data as in \cite{goffman,cron}.

\begin{figure}
\begin{center}
\includegraphics[width=8.5cm]{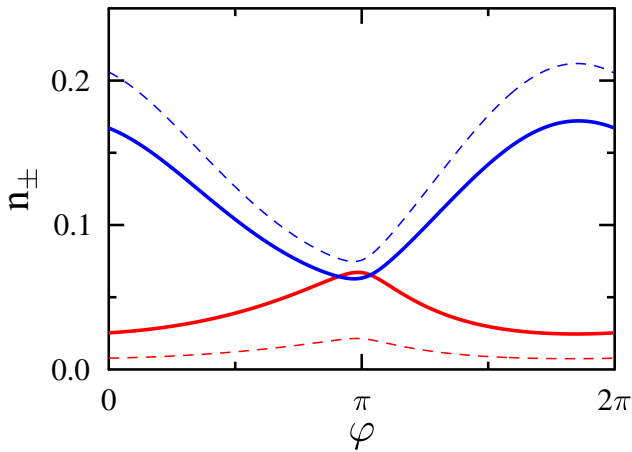}
\end{center}
\vspace*{-0.5cm}
\caption{\label{fig:populations}Steady state populations for the  $n_-$ [upper solid (blue) line] and the  $n_+$ [lower solid (red) line] Andreev levels for a transmission $\tau=0.99$ according to (\ref{adiab_WSG}). Also shown are the populations according to the dynamics on uncoupled surfaces (dashed lines) \cite{goffman,cron}. Parameters are $k_{\rm B} T/\Delta_S= 0.8, \varphi_0 I_{\rm dc}/\Delta_S=0.3$, and $\hbar\gamma/\Delta_S = 11$.}
\end{figure}

\begin{figure}
\begin{center}
\includegraphics[width=8.5cm]{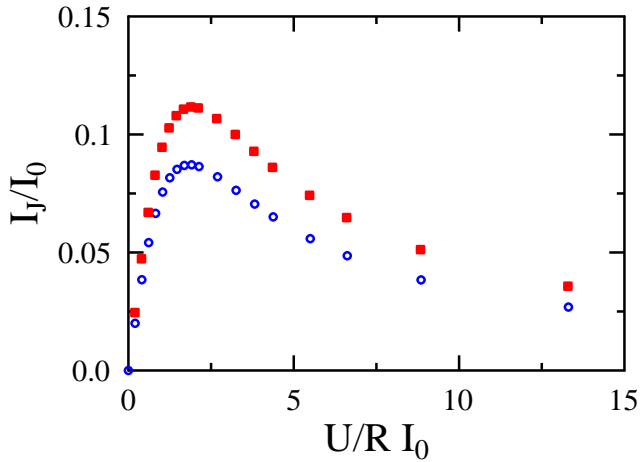}
\end{center}
\vspace*{-0.5cm}
\caption{\label{fig:ivcurves}Average supercurrent vs. average voltage in units of $R I_0$ with $I_0=(e\Delta_S/\hbar) (1-\sqrt{1-\tau})$ being the critical current of the junction. Parameters are the same as in fig.~\ref{fig:populations}. Squares (red) denote the results for the coupled dynamics, circles (blue) for the standard dynamics on averaged surfaces (see text).}
\end{figure}

\begin{figure}
\begin{center}
\includegraphics[width=8.5cm]{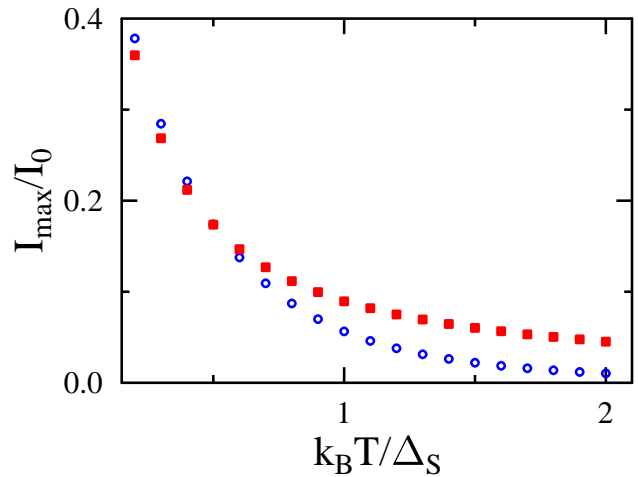}
\end{center}
\vspace*{-0.5cm}
\caption{\label{fig:supermax}Maximal supercurrent as a function of temperature. Parameters are chosen as in fig.~\ref{fig:populations}. Squares (red) depict the data for coupled dynamics, circles (blue) for the standard dynamics on averaged surfaces (see text).}
\end{figure}

We start with the transition rate (\ref{st_rate}) between the adiabatic surfaces and compare it with the relevant coupling frequency at the LZ point $\Omega_0=2\Delta_S\sqrt{1-\tau}/\hbar$. As long as $\Gamma/\Omega_0\ll 1$ the adiabatic approximation applies, while for $\Gamma/\Omega_0\simeq 1$ nonadiabatic transitions play a dominant role. Note that $\Gamma/\Omega_0$ diverges for $\tau\to 1$.
\begin{figure}
\begin{center}
\includegraphics[width=8.5cm]{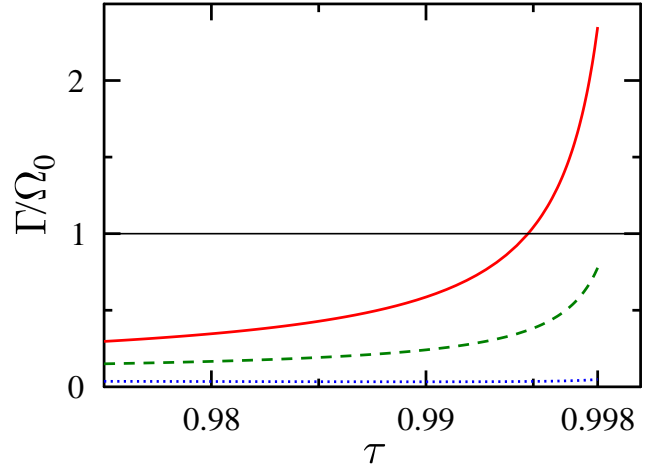}
\end{center}
\vspace*{-0.5cm}
\caption{\label{fig:transitionrates}Transition rate according to (\ref{st_rate}) and scaled with the frequency $\Omega_0=\Delta_0(\pi)/\hbar$ vs. the transmission for various inverse temperatures $\Delta_S/k_{\rm B} T=$ 1 (solid), 2 (dashed), 5 (dotted). Other parameters are $E_C/\Delta_S=2$, $\gamma\hbar/\Delta_S=11$, and $I_{\rm dc}\varphi_0/\Delta_S=0.3$. See text for details.}
\end{figure}
From fig.~\ref{fig:transitionrates} one observes that in the low temperature range the adiabatic approximation is well justified even for transmission very close to 1. The opposite is true for somewhat higher temperatures, where the ratio $\Gamma/\Omega_0$ becomes of order 1 or larger. In this domain nonadiabatic transitions must be taken into account and the standard Smoluchowski-type of phase diffusion on single Andreev levels breaks down. This can also be seen from the steady state populations depicted in fig.~\ref{fig:populations}. For high transmissions the populations around the LZ point differ substantially compared to those for isolated dynamics.
The supercurrent peak as a function of voltage is shown in fig.~\ref{fig:ivcurves}. For comparison results of the effective approach outlined in Sec.~\ref{adiabatic} are included as well, where a mean potential surface  consisting of a thermal average of the two Andreev levels is used in a standard Smoluchowski equation [see (\ref{meanpot}),(\ref{smoluch_strom})].
In accordance with the above discussion non-adiabatic transitions do increase the supercurrent $I_J$ compared to the case of the dynamics on averaged surfaces, because more (less) population is residing on the upper (lower) level in the down-hill direction.
Experimentally, the variance of the switching current (essentially the maximal supercurrent $I_{\rm max}$) with temperature has been found to deviate substantially for high transmitting channels from the standard Smoluchowski prediction \cite{goffman,cron}. In fact, with increasing temperature $I_{\rm max}$ has been observed to be {\em larger} than the predicted values, for very low temperatures smaller. The simple strategy to include non-adiabatic transitions by using the standard Landau-Zener formula mentioned in the Introduction, however, failed to capture this effect consistently. As illustrated in fig.~\ref{fig:supermax} the coupled dynamics does indeed describe the observed phenomena qualitatively. The increase in $I_{\rm max}$ for rising temperatures can be attributed to the larger transition rates seen in fig.~\ref{fig:transitionrates}. A quantitative comparison with the experimental data necessitates a more careful analysis of the actual circuit (see also below) and will be presented elsewhere.

\section{Role of quantum fluctuations}\label{quant}

The theory developed so far is basically a classical one. As discussed at the end of Sec.~\ref{overdamped} quantum effects in the momentum-momentum correlation can be effectively accounted for by working with the full correlation (\ref{cpp}) resp.\ its Fourier transform.  This seems  not to be consistent in a strict sense since quantum fluctuations in position are completely ignored. Here, we will provide arguments why this strategy is justified and to what extent the latter ones may appear in the theory.

The classical Smoluchowski theory requires not only strong friction $\gamma/\omega_0\gg 1$ but also $\gamma\hbar\beta\ll 1$. The opposite limit where the quantum scale for friction by far exceeds the thermal energy scale, i.e.\ $\gamma\hbar\beta\gg 1$, has been studied in \cite{lehle,smoluprl,springer}. The essence is this: In the so-called quantum Smoluchowski regime to leading order the classical Smoluchowski equation still applies. Quantum fluctuations appear in form of a modified diffusion coefficient
$k_{\rm B} T\to k_{\rm B} T/[1-\Lambda V''(q)/k_{\rm B} T]$ where $\Lambda=(\hbar/m\pi\gamma) \ln(\gamma\hbar\beta/2\pi)$ for $\gamma\hbar\beta\gg 1$. Typically, $\Lambda$ is small and describes deviations of equilibrium fluctuations in position from its classical value $\langle q^2\rangle_\beta-\langle q^2\rangle_{\beta,\rm cl}$. In contrast to this squeezing in position, fluctuations in momentum are large and fully quantum mechanical in agreement with the uncertainty principle. It is thus justified even for $\gamma\hbar\beta\gg 1$ to use at least to leading order the classical Smoluchowski equation for the dynamics in position, but the quantum version of the momentum-momentum correlation. To next order, quantum fluctuations in position are accounted for by the above replacement of the diffusion coefficient in (\ref{adiab_WSG}).

One may wonder in which regime, classical or quantum, superconducting point contacts are operated. For this purpose one realizes that $\gamma\hbar\beta=\beta E_C/\pi \rho\equiv\kappa$ where $E_C=2 e^2/C$ is the charging energy and $\rho=R/R_Q$ with $R_Q=h/4 e^2$ \cite{ankerjj}. In the overdamped limit one always has $\rho\ll 1$. For a contact with a capacitance in the fF range and $R$ of the order of 200$\Omega$ one then gets $\kappa\approx$ 40/T[K] so that even for temperatures of a few K, we have $\kappa\gg 1$. The conclusion is that the phase dynamics of the junction is taking place in the quantum Smoluchowski regime. We note that quantum fluctuations correspond here to charging effects and display Coulomb blockade physics \cite{ankerjj}.
Question is why this has not been observed yet. There are two answers. The first one is based on the above argument that even in this regime to leading order everything is classical. The second one is based on an analysis of the real circuitry. There, additional capacitances $C_s$ of typically a few pF are placed in parallel to the weak link meaning that it sees a more complex admittance with additional voltage fluctuations. These latter ones are classical since $\hbar\beta/R C_s\ll 1$ and may mask the quantum fluctuations. If either or both of these arguments apply needs a more careful study of the experimental situation and goes beyond the scope of the present work. Anyway,  atomic point contacts may be ideal test-beds to tune an overdamped system between its classical and quantum regimes.

\section{Summary}

We developed a consistent approach to describe in the regime of strong friction non-adiabatic transitions between  adiabatic surfaces. This results in a set of coupled equations of motion for the adiabatic populations (\ref{adiab_WSG}), which may be of use in a broad range of contexts. An explicit expression for the transition rate reveals that the strength of these transition is controlled by the inverse of a LZ-type of factor measuring the diabatic forces at the LZ-point relative to the energy gap between the adiabatic levels and the momentum-momentum correlation. Explicit results are obtained for high transmitting superconducting atomic point contacts. In particular, the experimentally observed temperature dependence of the maximal supercurrent is qualitatively explained. The classical theory may be extended to include quantum fluctuations.

\section*{Acknowledgements}
Financial support has been provided by the DFG through SPP1243 and the Landesstiftung BW through the network ''Functional Nanostructures III".

\section*{Appendix}
Here, we give a brief account on how to obtain the propagator for the off-diagonal elements in (\ref{adia_WFP}) in the overdamped limit. In this domain and in the time window $ 1/\gamma\ll t\ll\gamma/\omega_0^2$ ($\omega_0$ is a typical frequency of the system) the dynamics of the propagator ${\cal G}(q,p;\bar{q},\bar{p};t)$ according to $d{\cal G}/dt={\cal L}_0 {\cal G}$ is determined by only local properties of the potential $V_+$. Thus, we may use a local harmonic approximation $V_+(q)'\approx V_+'(\bar{q})+V_+''(\bar{q})(q-\bar{q})$. With $\Omega^2=V_+''(\bar{q})/m^2$ and $f=V'_+(\bar{q})$ the operator ${\cal L}_0$ in (\ref{lop}) takes the form
\begin{eqnarray}
	{\cal L}_0 &=& -\frac{p}{m} \frac{\partial}{\partial q}+ (f+m \Omega^2 q) \frac{\partial }{\partial p}\nonumber\\
				&&
				+ \gamma \left( \frac{\partial}{\partial p} p
						+ m k_B T \frac{\partial^2 }{\partial p^2} \right)\, .
		\label{def_greensfkt}
\end{eqnarray}
The propagator for this harmonic problem with the initial condition ${\cal G}(q,p;\bar{q},\bar{p},0)=\delta(q-\bar{q})\delta(p-\bar{p})$ is well-known
\begin{eqnarray}
\label{greensfunktion}
\lefteqn{{\cal G}(q,p;\bar{q}, \bar{p};t) = }\nonumber\\
&&\frac{1}{2 \pi \sqrt{F G -H^2}}\exp \left[- \frac{G R^2 - 2 HR Q + FQ^2}{2 (FG - H^2)}\right]\, ,
\end{eqnarray}
where the full time dependence is captured by the correlation functions
\begin{eqnarray}
	F &=& \frac{ k_{\rm B} T}{m\Omega^2} \left\{ 1
				- {\rm e}^{-\gamma t}
				\left[ 2 (\gamma^2/\nu^2) \sinh^2({\nu t}/{2})\right.\right.\nonumber\\
				 &&\left.\left.+ (\gamma/\nu) \sinh (\nu t) + 1 \right]\right\}\, ,
				\nonumber \\
	G &=&  m  k_{\rm B} T \left\{ 1
				- {\rm e}^{- \gamma t}
				\left[ 2 (\gamma^2/\nu^2) \sinh^2({\nu t}/{2})\right.\right.\nonumber\\
				&&\left.\left.- (\gamma/\nu) \sinh(\nu t) + 1\right]\right\}\, ,
				\nonumber \\
	H &=& (4 \gamma k_{\rm B} T/\nu^2) {\rm e}^{- \gamma t} \sinh^2(\nu t/2)\, ,
\end{eqnarray}
and
\begin{equation}
R= q - q_R(t)\ \ , \ \ Q = p - p_Q(t)\, .
\end{equation}
Here, $\nu = \sqrt{\gamma^2 - 4\Omega^2}$ and
 the latter two functions carry the mean dynamics of position and momentum, i.e,
\begin{eqnarray}
	q_R(t) &=& {\rm e}^{-\gamma t/2}\left[\left(\bar{q}+\frac{f}{m\Omega^2}\right)  \cosh\left( \frac{\nu t}{2}\right)\right.\nonumber\\
					&&\left.+ \frac{(\bar{q}+f/m\Omega^2)\gamma + 2 \bar{p}/m}{\nu} \sinh \left(\frac{\nu t}{2} \right) \right]-\frac{f}{m\Omega^2}\nonumber\\
	p_Q(t) &=& \bar{p}e^{-\gamma t/2} \cosh \left( \frac{\nu t}{2} \right)
					- \frac{2 (f+\bar{q} m \Omega^2) + \gamma \bar{p}}{\nu} e^{- \gamma t/2}\nonumber\\
					&&\times\sinh \left( \frac{\nu t}{2} \right).					
	\label{q_p_Lsg}
\end{eqnarray}
Now, within the time window  $1/\gamma\ll t\ll\gamma/\Omega^2$  the correlation $H$ becomes of order $1/\gamma$, while the product $F G$ is larger than $H^2$ by a factor $\gamma t\gg 1$. Accordingly, one shows that the $q$ and the $p$-dependence in ${\cal G}$ factorize. Further, for the small parameter $\epsilon=t/(\gamma/\Omega^2)\ll 1$ the $q$ dependence becomes a Gaussian sharply peaked around $q=\bar{q}$ with a width of order $\epsilon$ thus representing effectively a $\delta(q-\bar{q})$ contribution. In contrast, the $p-\bar{p}$ fluctuations are of  order 1. This way, one arrives at the result specified in (\ref{greensfkt_approx}). As expected the dynamics in position is frozen within the time window, while the momentum equilibrates to its instantaneous value around $p_Q(t)$.


\begin{thebibliography}{99}

\bibitem{landau} L. D. Landau, Phys. Z. Sowjetunion {\bf 1}, 89 (1932).

\bibitem{zener}C. Zener, Proc. R. Soc. London {\bf A 137}, 696 (1932).
	
\bibitem{stuckel}E.G.C. St\"uckelberg, Helv. Phys. Acta {\bf 5}, 369 (1932).
	
\bibitem{nitzan}A. Nitzan, Annu. Rev. Phys. Chem. {\bf 52}, 681 (2001).
	
\bibitem{garg}A. Garg, J. N. Onuchic, and V. Ambegaokar, J. Chem. Phys. {\bf 83}, 4491 (1985).
	
\bibitem{hanggi}L. Hartmann, I. Goychuk, and P. H{\"a}nggi, J. Chem. Phys. {\bf 113}, 11159 (2000).
	
\bibitem{pollak}M.-L. Zhang, S. Zhang, and E. Pollak, J. Chem. Phys. {\bf 119}, 11864 (2003).

\bibitem{lehle}H. Lehle and J. Ankerhold, J. Chem. Phys. {\bf 120},1436 (2003).
	
\bibitem{scheer}E. Scheer, P. Joyez, D. Esteve, C. Urbina, and M.H. Devoret, Phys. Rev. Lett. {\bf 78}, 3535 (1997).
	
\bibitem{scheer2}E. Scheer, N. Agra{\"{i}}t, J.C. Cuevas, A. Levy Yeyati, B. Ludoph, A. Mart{\'i}n-Rodero, G. Rubio Bollinger, J.M. van Ruitenbeek, and C. Urbina, Nature {\bf 394}, 154 (1998).
	
\bibitem{yeyati}N. Agra{\"{\i}}t, A. Yeyati, and J. van Ruitenbeek, Phys. Rep. {\bf 377}, 81 (2003).
	
\bibitem{barone}A. Barone, {\em Physics and applications of the Josephson effect}, (John Wiley \& Sons, 1982).
			
\bibitem{beenakker}C.W.J. Beenakker, Phys. Rev. Lett. {\bf 67}, 3836 (1991).
	
\bibitem{tinkham}T. M. Klapwijk, G.E. Blonder, and M. Tinkham, Physica B+C {\bf 110}, 1657 (1982).
	
\bibitem{averin}D. Averin and A. Bardas, Phys. Rev. Lett. {\bf 75}, 1831 (1995).
	
\bibitem{cuevas}J.C. Cuevas, A. Mart{\'i}n-Rodero, and A. Levy Yeyati, Phys. Rev. B {\bf 54}, 7366 (1996).
	
\bibitem{chauvin}M. Chauvin, P. vom Stein, D. Esteve, C. Urbina, J.C. Cuevas, and A. Levy Yeyati, Phys. Rev. Lett. {\bf 99}, 067008 (2007).
	
\bibitem{halperin}V. Ambegaokar and B.I. Halperin, Phys. Rev. Lett. {\bf 22}, 1364 (1969).
	
\bibitem{ivanchenko}Yu. M. Ivanchenk and L.A. Zil'berman, Sov. Phys. JETP {\bf 28}, 1272 (1969).
	
\bibitem{yeyati2}R. Duprat and A. Levy Yeyati, Phys. Rev. B {\bf 71}, 054510 (2005).
	
\bibitem{goffman}M.F. Goffman, R. Cron, A. Levy Yeyati, P. Joyez, M.H. Devoret, D. Esteve, and C. Urbina, Phys. Rev. Lett. {\bf 85}, 170 (2000).
	
\bibitem{cron}R. Cron, {\em Atomic contacts: a test-bed for mesoscopic physics}, (CEA Saclay, 2001).
	
\bibitem{ivanov}D. A. Ivanov and M. V. Feigel'man, Phys. Rev. B {\bf 59}, 8444 (1999).
	
\bibitem{weiss}U. Weiss, {\em Quantum Dissipative Systems}, (World Scientific, 2003).
			
\bibitem{breuer}H.-P. Breuer and F. Petruccione, {\em The theory of open quantum systems}, (Oxford University Press, 2000).
			
\bibitem{zusman}L. D. Zusman, Chem. Phys. {\bf 49}, 295 (1980).
	
\bibitem{remark}Note that in the ohmic case the frequency integral must be supplemented by a high frequency cut-off to be convergent. Since here only the Fourier transform is needed, we do not need to specify it explicitly.

\bibitem{smoluprl}J. Ankerhold, P. Pechukas, and H. Grabert, Phys. Rev. Lett. {\bf 87}, 086802 (2001); J. Ankerhold and H. Grabert, {\em ibid} {\bf 101} (E), 119903 (2008).

\bibitem{springer}J. Ankerhold, {\em Quantum Tunneling in Complex Systems}, STMP {\bf 224}, (Springer, 2007).
			
\bibitem{ankerjj}J. Ankerhold, Europhys. Lett. {\bf 67}, 280 (2004).
	
\end{thebibliography}
\end{document}